\documentclass{iau} 
\usepackage{graphicx}
\usepackage{amsmath}
\usepackage{hyperref}
\usepackage[percent]{overpic}
\hypersetup{
    colorlinks=true,
    linkcolor=blue,
    filecolor=magenta,      
    urlcolor=cyan,
}
 
\title[S stars in the Gaia era] 
{Probing stellar evolution with S stars and Gaia}

\author[S. Shetye et al]   
{S. Shetye$^1$$^,$$^2$ \thanks{This work has made use of data from the European Space Agency (ESA)
mission \textit{Gaia} (\url{https://www.cosmos.esa.int/gaia}), processed by
the \textit{Gaia} Data Processing and Analysis Consortium (DPAC,
\url{https://www.cosmos.esa.int/web/gaia/dpac/consortium}). Funding for
the DPAC has been provided by national institutions, in particular the
institutions participating in the \textit{Gaia} Multilateral Agreement.}, S. Van Eck$^1$, A. Jorissen$^1$, H. Van Winckel$^2$, L. Siess$^1$ and S. Goriely$^1$}

\affiliation{$^1$Institut  d'Astronomie  et  d'Astrophysique,  Universit\'e  Libre  de  Bruxelles,  CP  226,  Boulevard  du  Triomphe,  B-1050  Bruxelles,Belgium \\ email: {\tt Shreeya.Shetye@ulb.ac.be} \\[\affilskip]
$^2$Instituut voor Sterrenkunde (IvS), KU Leuven, Celestijnenlaan 200D, B-3001 Leuven, Belgium }

\pubyear{2018}
\volume{343}  
\jname{Why galaxies care about AGB stars?}
\editors{F. Kerschbaum, M. Groenewegen, \& H. Olofsson, eds.}
\begin{document}

\maketitle

\begin{abstract}
S-type stars are late-type giants enhanced with s-process elements originating either from nucleosynthesis during the Asymptotic Giant Branch (AGB) or from a pollution by a binary companion. The former are called intrinsic S stars, and the latter extrinsic S stars. The intrinsic S stars are on the AGB and have undergone third dredge-up events. The atmospheric parameters of S stars are more numerous than those of M-type
giants (C/O ratio and s-process abundances affect the thermal structure and spectral synthesis), and
hence they are more difficult to derive. These atmospheric parameters are also entangled within each other. Nevertheless, high-resolution spectroscopic data of S stars combined with the Gaia Data Release~2~(GDR2) parallaxes and with the MARCS model atmospheres for S-type stars were used to
derive effective temperatures, surface gravities, and luminosities. These parameters not only allow to locate the intrinsic and extrinsic S stars in the Hertzsprung-Russell (HR) diagram but also allow the accurate abundance analysis of the s-process elements.  
\end{abstract}

\firstsection 
\section{Introduction}

S stars are late-type giants showing ZrO molecular bands along with TiO bands as the most characteristic distinctive spectral features (\cite{Merrill}). The C/O ratio of S stars ranges from 0.5 to 1  suggesting that they are transition objects between M-type giants (C/O $\sim  0.5$) and carbon stars (C/O $>$ 1) on the Asymptotic Giant Branch (AGB) (\cite{IbenRenzini}). Their spectra show signatures of overabundances in s-process elements (\cite{smithandlambert}).

The evolutionary status of S stars as AGB stars was challenged when Tc lines (a s-process element with no stable long-lived isotope) were reported as missing in some S stars (\cite{merrill1952}; \cite{smithandlambert1986}; \cite{alain1993}). This puzzle of the evolutionary status of S stars was solved when it was perceived that the Tc-poor S stars belong to binary systems (\cite{smithandlambert1986}; \cite{alain1993}). S stars can therefore be classified into two different classes: Tc-rich as intrinsic S stars that are genuine thermally-pulsing AGB (TP-AGB) stars  and Tc-poor as extrinsic S stars that correspond to the cooler analogues of barium stars and owe their s-process element overabundances to a mass transfer from a former AGB companion which is now a white dwarf.

The thermal structure of the atmospheres of S stars depend on the effective temperature (T\textsubscript{eff}), surface gravity (log g), [Fe/H], C/O as well as [s/Fe] (s-process element abundances). Abundance analysis of S stars thus requires a reliable determination of all these stellar atmosphere parameters.

\section{Stellar sample and parameter determination}

Our sample consists of S stars from the General Catalog of S stars (\cite{cgsscatalog}) with the condition to have a V-band magnitude brighter than 11 and $\delta \ge -30^{\circ}$, to be observable with HERMES (High Efficiency and Resolution Mercator Echelle Spectrograph, mounted on the 1.2m Mercator Telescope at the Roque de Los Muchachos Observatory, La Palma, \cite{raskinetal2011}). 
Furthermore, a condition was imposed on the TGAS parallaxes (\cite{gaia}), considering only those stars with a small error on the parallax ($\sigma_{\bar\omega} \leq 0.3 \bar\omega$). With these conditions, the sample amounts to 18 S stars. During the course of our study, the Gaia Data Release 2 (GDR2;  \cite{gdr2}) parallaxes were released and these more accurate parallaxes are used in the present study.

The stellar parameters were derived using the MARCS grid of atmospheric models for S stars (\cite{VanEck2017}) containing more than 3500 models covering the parameter space in T\textsubscript{eff}, log g, [Fe/H], C/O and [s/Fe] ratios. The comparison between observed and synthetic spectra is then performed by a $\chi^2$-fitting procedure, summing over all spectral pixels in spectral bands approximately 200 \textup{\AA} wide. The model with the lowest  $\chi^2$ value is chosen as the best fitting model. 

\begin{figure}
\begin{center}
\includegraphics[scale=0.25]{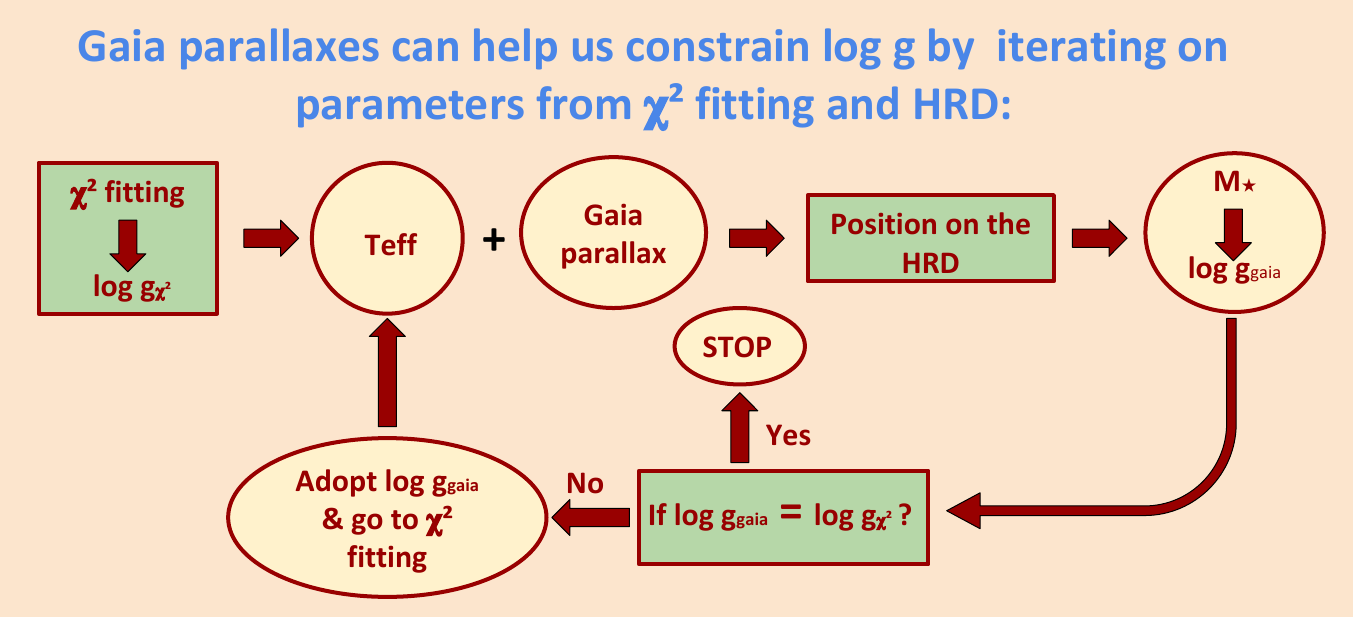}
\caption{\label{loggiterations}Algorithm adopted to constrain log g using the locations of the stars in the HR diagram compared to the evolutionary tracks from STAREVOL code. }
\end{center}
\end{figure}

The distance and luminosity of these stars was determined from the GDR2 parallaxes. Comparison between the positions of the stars in the HR diagram constructed from GDR2 parallaxes with the evolutionary tracks from the STAREVOL code (\cite{lsiess2008}) yield the mass, hence the surface gravity of our stars (log g \textsubscript{Gaia}). Because log g \textsubscript{Gaia} and log g derived from the $\chi^2$  fitting do not always agree, we derive a new surface gravity estimate as explained in Figure \ref{loggiterations}. This iteration on the stellar parameters ensures that the adopted log g is consistent with the GDR2 parallaxes.

\section{HR diagram of S stars}
\begin{figure*}      
\begin{centering}
    \mbox{\includegraphics[scale=0.32,trim={1.5cm 0cm 0cm 0cm}]{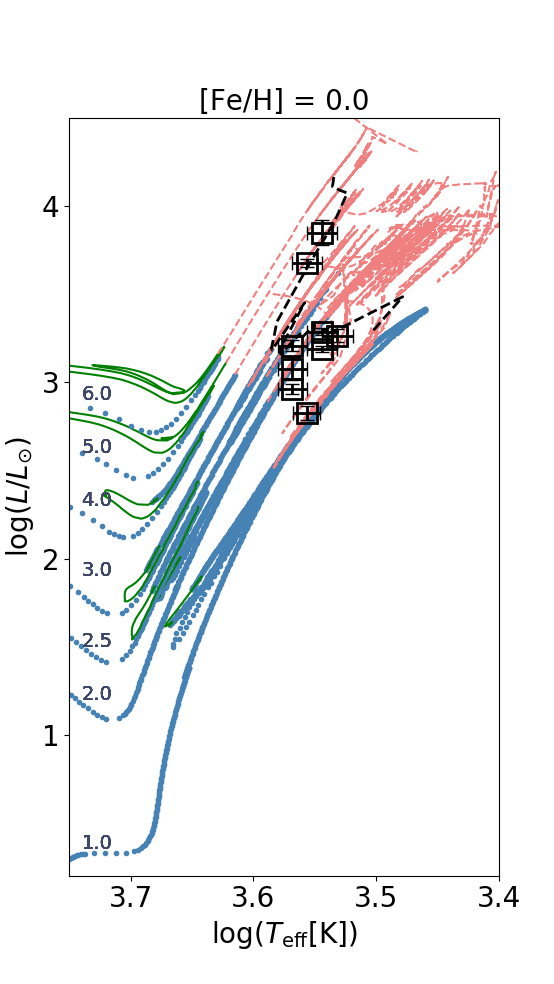}}   
    \hspace{0px}
    \begin{overpic}[scale=0.32,trim={1.3cm 0cm 0cm 0cm}]{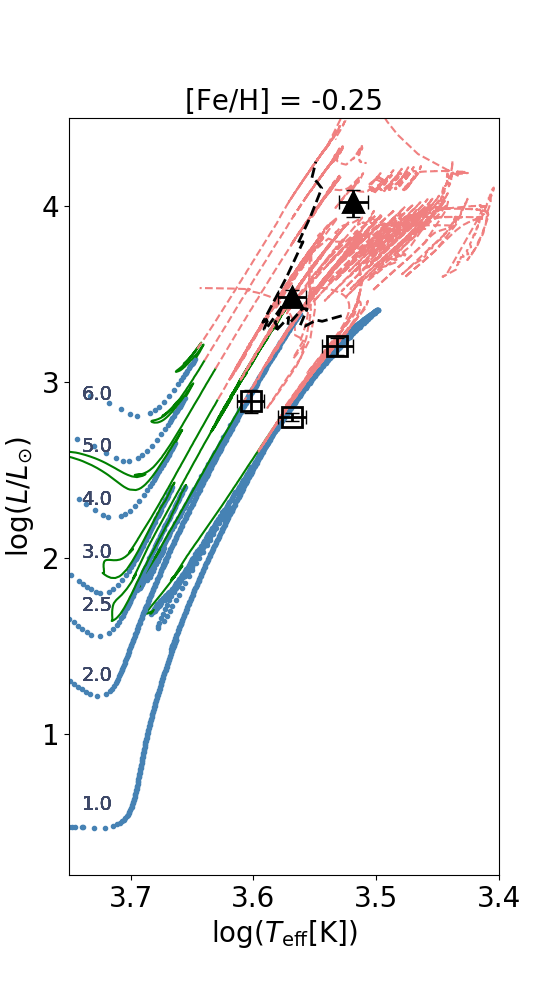}  
     \put(34,78){\textbf{\tiny{UY Cen}}}
     \put(11,68){\textbf{\tiny{NQ Pup}}}
    \end{overpic}
    \hspace{0px}
    \begin{overpic}[scale=0.32,trim={1.3cm 0cm 0cm 0cm}]{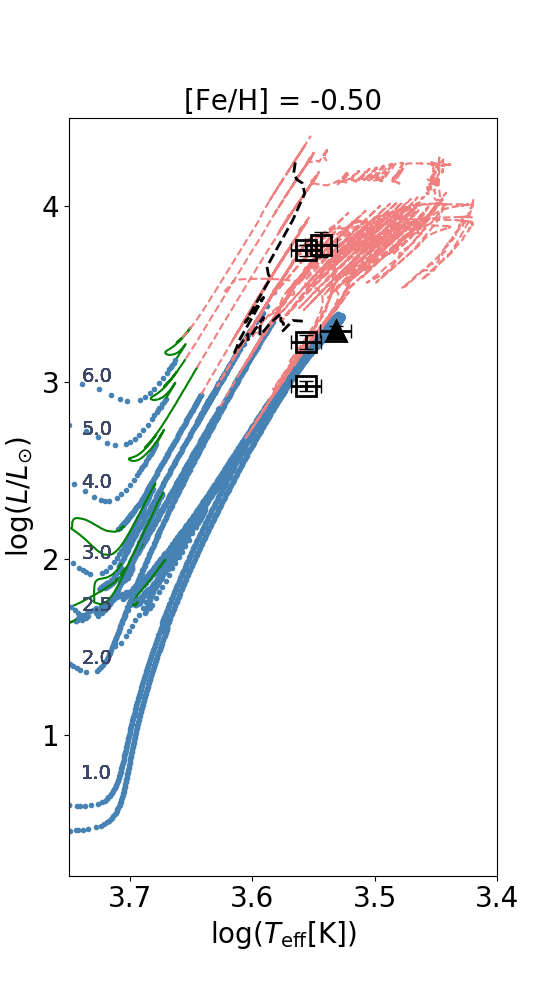}  
     \put(31,65){\textbf{\tiny{V915 Aql}}}
    \end{overpic}
    \caption{\label{HRD}HR diagram of intrinsic (filled triangles) and extrinsic (open squares) S stars along with the STAREVOL evolutionary tracks corresponding to the closest grid metallicity. 
    The red giant branch is represented by the blue dots, the core He-burning phase by the green solid line, whereas the red dashed line corresponds to the AGB tracks. The black dotted line marks the predicted onset of third dredge-up (TDU) from the STAREVOL code.}
    \label{materialflowChart}
\end{centering}
\end{figure*}

The final parameters and GDR2 parallaxes along with the STAREVOL evolutionary tracks corresponding
to the closest grid metallicity lead to the HR diagram of S stars presented in Figure \ref{HRD}. The intrinsic S stars are cool and luminous objects likely on the TP-AGB in the HR diagram. On the other hand, the extrinsic S stars are hotter and intrinsically fainter on the early AGB or more likely on the red giant branch (RGB). Interestingly, the Tc-poor S stars with M $\leq$ 2 M$_\odot$ are on the upper-end of the RGB or on the early-AGB while the Tc-poor S stars with M $\geq$ 2 M$_\odot$ are on the early- AGB. We also find a remarkable agreement between the location of the stars in the HRD and the predicted boundary for the occurence of TDU.

\section{A Tc-rich S star of M $\sim$ 1~M$_\odot$}

\begin{figure}
\begin{center}
\includegraphics[scale=0.6]{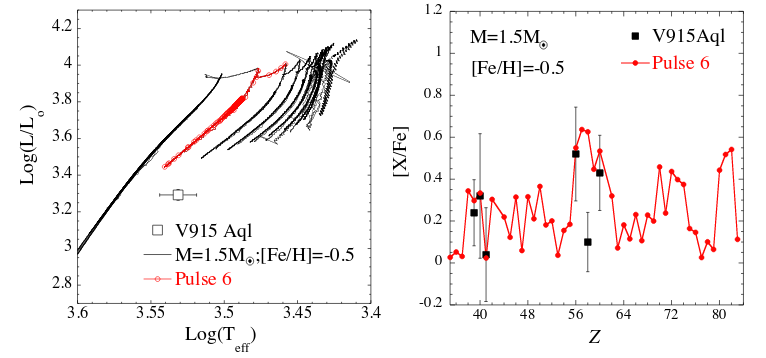}
\caption{\label{V915abund} Left panel: Location of the V915 Aql in the HR diagram, compared with STAREVOL tracks of the corresponding metallicity.
Right panel: Predicted abundance distribution for this object.}
\end{center}
\end{figure}

V915 Aql is an intrinsic S star located on the 1 M$_\odot$ track. It is intriguing because stellar models do not predict the third dredge-up to occur for masses less than 1.3 M$_\odot$ (\cite{karakas}). Figure \ref{V915abund} presents the comparison of the observed s-process elements abundance with nucleosynthesis predictions from the STAREVOL code (\cite{stephane}). To match V915 Aql surface abundances, a 1.5 M$_\odot$ model had to be used, but such a model does not match so well its position in the HR diagram. Nevertheless, abundance predictions for this model reproduce fairly well the overall pattern of V915 Aql (except for La). The occurence of the third dredge-up for low mass stars ( $<$ 1.3 M$_\odot$) was also found in low-luminosity s-process rich post-AGB stars (\cite[De Smedt et al. 2015]{kenneth}).

\section{Conclusion}
Combining  the  derived  parameters  with GDR2 parallaxes  allows  a  joint  analysis  of  the  location  of  the  stars  in  the Hertzsprung-Russell diagram and of their surface abundances. The Tc-rich star V915 Aql is challenging as it points at the occurrence of TDU episodes in stars with masses as low as M $\sim$ 1 M$_\odot$. Extending the sample to include many more intrinsic S stars with GDR2 parallaxes in order to constrain
the luminosity of the first occurrence of the TDU is a work in progress.

\begin{discussion}

\discuss{Uttenthaler}{You said that intrinsic S-type stars are the first stars on the AGB to show signs of 3DUP. However, there are pure M-type AGB stars that show one sign of 3DUP, namely lines of Tc; they have no ZrO bands. Would these objects be interesting for you to study and do you have plans to do so?}

\discuss{Shetye}{Yes, these objects will be indeed very interesting to study. It will be interesting to compare the positions of these stars with the intrinsic (Tc-rich) S-stars in the HR diagram. Thanks for the suggestion, it will definitely be added to our future plans.}

\end{discussion}

\end{document}